\documentclass[global,referee]{svjour}
\usepackage{subfigure}
\usepackage[dvips]{graphicx}
\usepackage{float}
\journalname{Ocean Dynamics}
\begin{document}
\title{Comparison of reduced-order,  sequential and variational
data assimilation methods in the tropical Pacific Ocean}
\author{C\'eline Robert\inst{1} \thanks{\emph{Present
address:} Celine.Robert@imag.fr}
\and Eric Blayo\inst{1} \and Jacques Verron \inst{2}%
}                     
%
%
\institute{LMC-IMAG UMR 5523 CNRS/INPG/UJF/INRIA, Grenoble \and LEGI UMR 5519
CNRS/INPG/UJF, Grenoble}
\date{Received: date / Revised version: date}
\titlerunning{Comparison of reduced-order data assimilation methods}
\maketitle
\textbf{\today}
\begin{abstract}\par
This paper presents a comparison of two reduced-order, sequential and variational data assimilation
methods: the SEEK filter and the R-4D-Var. A hybridization of the two, combining the variational 
framework and the sequential evolution of covariance matrices,  is also preliminarily investigated 
and assessed in the same experimental conditions.
The comparison is performed using the twin-experiment approach on a model of the Tropical Pacific domain. 
The assimilated data are simulated temperature profiles at the
locations of the TAO/TRITON array moorings.
It is shown that, in a quasi-linear regime, both methods produce similarly good results. 
However the hybrid approach provides slightly better results and thus appears as potentially fruitful. 
In a more non-linear regime, when Tropical Instability Waves develop, the global nature 
of the variational approach helps control model dynamics better than the sequential 
approach of the SEEK filter. This aspect is  probably enhanced by the context of the experiments 
in that there is a limited amount of assimilated data and no model error.
\end{abstract}

\newcommand{\bs}[1]{\boldsymbol{#1}} 
\newcommand{\tbf}[1]{\textbf{#1}}
\newcommand{\nino}{El Ni\~no} 

\newcommand{\ds}{\displaystyle}

\section{Introduction}
Operational oceanography is an emerging field of activity that is concerned with  real-time monitoring and prediction of the physical and biogeochemical state of oceans and regional seas. Operational ocean prediction systems have been made feasible by the concomitance of several elements: the emergence of relatively reliable numerical models and of appropriate computing capabilities, the establishment of global ocean observation systems, and the progress achieved in data assimilation techniques. It is the latter of these advances that is addressed in this paper.
In the geophysical context, data assimilation methods  face a number of specific difficulties. In particular, due to the very large dimensions of the systems, the computational burden and the prescription of adequate error statistics are critical issues. In addition, there is a need to improve methods in the case of  non-linear systems and/or non-gaussian statistics.
\par
Data assimilation methods are generally classified into two groups according to the approach used: the sequential approach, based 
on the statistical estimation theory and the Kalman filter, and the variational approach (4D-Var), built from the optimal control theory.
It is well known that the 4D-Var and Kalman filter approaches provide the same solution, at the end of the assimilation period, 
for perfect and linear models. These approaches are different however, mainly because the model is seen as a strong constraint 
in the 4D-Var approach and as a weak constraint in the sequential approach. In addition, the specification and time evolution of 
the error statistics,  the length and structure of the forecast-analysis cycles, and the temporal use of observations may be quite 
different. 
In practice, due especially to  non-linearity,  these differences can result in significant discrepancies between 
the solutions provided by the two approaches. \par
 The full Kalman filter cannot be used in actual geophysical systems, 
because specifying of the error covariance matrices
is difficult
 and also involves huge computational costs and impractical matrix handling.  The need to circumvent these difficulties 
 has led to the development of reduced-order Kalman filters.
 Here, order reduction consists in reducing the size of the background error covariance matrix
 by selecting a number of directions in the state space along which the error variability is assumed to lie.
 In recent years,
 this approach has given birth in particular to the Ensemble Kalman Filter (EnKF) (Evensen, 1994), the
 Reduced-Rank-SQuare-RooT (RRSQRT) filter (Verlaan and Heemink, 1997), the Singular Evolutive Extended Kalman (SEEK) filter (Pham \textit{et al.}, 1998, Verron \textit{et al.}, 1999) and the ESSE method (Lermusiaux and Robinson, 1999).
 These four methods basically differ in their strategies to approximate the error covariance matrix
 and/or the way in which they propagate the state error statistics. In the EnKF, the error statistics are propagated using a statistically relevant
ensemble of states. The forecast error covariance matrix is not given explicitly.
The SEEK and RRSQRT filters are based on a truncation of an eigendecomposition of the error
covariance matrix, and partly differ in their initial choice of the approximate low-rank matrix, and with respect to its time evolution. The SEEK takes advantage of the fact that the ocean is
a dynamic system with an attractor, and is not intended to make corrections
in directions perpendicular to the attractor, which are naturally attenuated by the system.
In this paper, the SEEK filter is chosen.  \par
 The variational 4D-Var method has long been used in
meteorology  (\textit{e.g.} Rabier, 1998) and has been applied to several operational forecasting
  systems in its incremental form (Courtier \textit{et al.}, 1994), a form that is particularly
   suited to nonlinear systems. It has also been developed for oceanographic situations
    (\textit{e.g.} Greiner and Arnault, 1998a, b; Vialard \textit{et al.}, 2002, 2003; Weaver \textit{et al.}, 2003).
     The method is costly and involves complex software development for the tangent linear and adjoint models.
As with sequential estimation, lack of knowledge concerning the error statistics leads
 to the use of approximations and models for the background error covariance matrix.
To solve the problem, the order reduction procedure can also be used for the 4D-Var
to build the Reduced-4D-Var (Blayo \textit{et
al.}, 1998). This has been tested in a realistic configuration by 
Robert \textit{et al.} (2005).
In the Reduced 4D-Var, the control parameter (namely the initial condition) now belongs
 to a low-dimension space and the background error covariance matrix can thus also be expressed 
 using this subspace. The Incremental Reduced 4D-Var (hereafter R-4D-Var) is therefore the variational 
 approach chosen here. \par
 A major advantage of the 4D-Var assimilation 
 is the simultaneous and consistent use of the whole observational dataset over
 the assimilation time window and the optimisation of the model trajectory is thus based on the global 
 processing of these observations.
 A serious drawback however is that the background error statistics are often constant over
 this assimilation time window  in actual applications.
 These characteristics are somewhat reversed with the Kalman filter:  observations are processed 
 sequentially, and are often grouped in actual applications (which means that they are not generally 
 processed at the exact time of observation),
 but the state error statistics can be propagated from one assimilation cycle to the next.
 To benefit from the advantages of each of both approaches,
 Veers\'e \textit{et al.} (2000)
 proposed a hybrid algorithm that combines the analysis performed by the variational
 approach with the state error propagation of the SEEK filter. This hybrid approach  is 
 the third type of data assimilation algorithm that is studied here.
\par
The main objective of this study is to compare these three types of 
reduced-order data assimilation approaches.
Note that in the reduced-order framework,
the error subspaces can be built using the same method (the choice of the basis can be the same) 
and the initial subspaces will be identical. The comparison is conducted using the twin-experiment approach, in which the data assimilated are synthetic and obtained from a free run of the model. 
Data and model are therefore entirely consistent and the assimilation is artificially facilitated 
as far as the model error  and data error characteristics are concerned. Given the current levels of observation, this approach was considered to be the only possible way to conduct a methodological comparison since the ocean is not 
sufficiently well observed for a real comparison exercise to be meaningful.  This will clearly 
be the next step following the present work.\par
To our knowledge, the present study represents is one of the first attemps to compare these two  methods  using exactly
the same configuration.  Since the model used in this study is (weakly) 
non-linear and the system is of large dimension, 
the results obtained by both methods should not be expected to be the same. 
The region chosen for the experiments is the tropical Pacific Ocean. The large-scale ocean dynamics in this region
 is weakly non-linear except for the Tropical
Instability Waves (TIWs) that develop in the eastern Pacific and
propagate along the equator, becoming increasingly intense from mid-June/early July.
The tropical Pacific Ocean was chosen because it is one of the best-observed regions of 
the world ocean thanks to the TOGA program and the TAO mooring network in particular. In addition, many
numerical studies have been performed in this area and, as a result, 
direct, tangent linear and adjoint models
have been reasonably well validated. \par
The article is organized as follows: in the next section (Section
\ref{Sect:methods}) we detail the methods used, introducing in particular the hybrid approach.
Then, we describe the configuration of the twin-experiment framework
(Section \ref{Sect:config-exps}). Finally, 
we present the main results obtained in each case (Section
\ref{Sect:results}), followed by
conclusions and discussion (Section \ref{Sect:conclu}).
%
%
\section{The reduced-order approach}
\label{Sect:methods}
This section provides details of the different reduced-order methods used in our experiments. 
In the following,  the notations proposed by Ide \textit{et al.},
(1997) are used. The superscripts $a$, 
$b$, $f$ and $T$ represent respectively the
analysis, background, forecast and mathematical transpose sign.\par
In the full dimension space, error covariances are unknown and must be
modeled. However, this is challenging for complex oceanic systems
since the state vector contains several physical quantities (velocity, temperature, salinity) and is very  large (it commonly reaches $n=10^6$ components), and because many different spatial scales interact. 
One way to try to overcome these difficulties is to consider that most of the variance can be retained within a low-dimension space,
spanned by a basis of a limited number of vectors.  The error covariance matrix can then be approximated by a low-rank matrix, considering only this reduced space.
To make the reduction computationally efficient, the number of retained vectors $r$ must be small
with respect to the number of degrees of freedom of the system ($r\ll n$). To make the data assimilation effective, however, the subspace must adequately represent the main directions of error propagation in the system. 
This type of order reduction is used in both the sequential method (SEEK
filter) and the variational approach (R-4D-Var).
\subsection{Definition of the subspace}
\label{Sect:methods-subspace}
In our experiment,  an EOF basis  is chosen to span the error subspace, and  will be used both for the SEEK filter and R-4D-Var implementations. This means that we assume that the variability of the model state vector is representative of the variability of the background error, which is indeed verified in the present context of twin experiments (\textit{i.e.}  with no model error).
Other bases can however be thought of for building this subspace, such as Lyapunov, singular or breeding  vectors (Durbiano, 2001), but EOFs have proved to be efficient in the present context, probably because they take into account the nonlinearity of the model dynamics, and also because their covariance matrix is relatively well known. \par
The model solution, obtained from a previous numerical simulation, is sampled
 and a multivariate EOF analysis of the resulting $p$ three-dimensional state vectors $({\bf x}(t_1),\ldots,{\bf x}(t_p))$ is performed.   It should be remembered
that this analysis aims at determining the main directions of variability of the model sample, which 
leads to diagonalizing  the empirical covariance matrix ${\bf X} {\bf X}^T$, where ${\bf X}=({\bf X}_1,\ldots,{\bf X}_p)$, with ${\bf X}_j(i)= \displaystyle \frac{1}{\sigma_i} [{\bf x}(t_j)-\bar{\bf x}]$, $\bar{\bf
x}=\displaystyle \frac{1}{p}\, \displaystyle \sum_{j=1}^{p}{\bf x}(t_j)$ and  ${\sigma_i^2}$ is the empirical variance of the $i$-th
component of the state vector: ${\sigma_i^2}= \displaystyle \frac{1}{p}\,\displaystyle \sum_{j=1}^{p}({\bf X}_j(i) )^2$.
The inner product is 
the usual one for a state vector containing several physical quantities expressed in different units: 
\begin{equation}
<{\bf X}_j,{\bf X}_k> = \sum_{i=1}^{n} \displaystyle \frac{1}{\sigma_i ^2} 
({\bf x}(t_j)-\bar{\bf x})_i ({\bf x}(t_k)-\bar{\bf x})_i
\end{equation}
Since the size $p$ of the sample is generally much smaller than the size $n$ of the model state vector, the actual diagonalization is performed on the $p\times p$ matrix ${\bf X}^T {\bf X}$ rather than on the  $n\times n$ matrix ${\bf X} {\bf X}^T$ (it is well known that those two matrices have the same spectrum).
This diagonalization leads to a set of orthonormal eigenvectors $({\bf
L}_1,\ldots, {\bf L}_p)$ corresponding to eigenvalues $\lambda _1 >
\ldots > \lambda_p >0$. Since trajectories are computed with
the free model, these modes represent its variability over the whole sampled period.\par
If the background error $e_B$ is modeled as spanned by the $r$ first EOFs: $\displaystyle{ e_B = \sum_{j=1}^r w_j {\bf L}_j = {\bf L} {\bf w} }$, then its covariance matrix is modeled by ${\bf B}_r = E(   {\bf L} {\bf w}  {\bf w}^T {\bf L}^T ) =  {\bf L}\, E( {\bf w}  {\bf w}^T ) {\bf L}^T$, which is approximated by ${\bf B}_r = {\bf L}  {\bf \Lambda}_r {\bf L}^T$ with ${\bf \Lambda}_r = \hbox{diag}(\lambda_1,\ldots, \lambda_r)$, since $\lambda_j$ is  the natural estimate for the covariance of ${\bf w}_j$.
The fraction of variability (or ``inertia") which is conserved when
retaining only the $r$ first vectors is $\displaystyle \sum_{j=1}^{r}
\lambda_j / \sum_{j=1}^{p} \lambda_j$. 
\subsection{The sequential approach}
\label{Sect:methods-seek}
In the sequential approach,   the SEEK filter is used following
Pham \textit{et al.} (1998) and Verron \textit{et al.} (1999).
Each error covariance matrix is decomposed in a reduced space in the form:
\begin{equation}
\textbf{P} = \textbf{S}\textbf{S}^{T}
\label{Eq:decompinit}
\end{equation}
The first estimate of the forecast error covariance matrix $\textbf{P}^f_0=\textbf{S}_0\textbf{S}_0^{T}$
 is given by the EOF decomposition, and can thus be written as 
 $\textbf{P}^f_0= {\bf L}  {\bf \Lambda}_r {\bf L}^T$.\par
The SEEK filter algorithm is composed of successive analysis-forecast
cycles. The analysis, at cycle $k$, is given by:
\begin{equation}
\textbf{x}^a_k = \textbf{x}^f_k + \textbf{K}_k [\textbf{y}_k - \textbf{H}_k \textbf{x}^f_k ]
\label{Eq:seek-analysis}
\end{equation}
${\bf K}_k$ is the gain matrix, which minimizes the variance of the
analysis error and thus satisfies the following equation 
(given that $\textbf{P}^f_k = \textbf{S}^f_k\textbf{S}^{fT}_k$):
\begin{eqnarray}
\textbf{K}_k   &=& \textbf{S}^f_k [{\bf I} + (\textbf{H}_k \textbf{S}^f_k)^T 
\textbf{R}^{-1}_k (\textbf{H}_k \textbf{S}^f_k)]^{-1} (\textbf{H}_k \textbf{S}^f_k)^T \textbf{R}^{-1}_k
\label{Eq:calculK}
\end{eqnarray}
The forecast, from cycle $k$ to $k+1$, is obtained using the 
model: 
\begin{equation}
\textbf{x}^f_{k+1} = M_{k,k+1} [\textbf{x}^a_k]
\label{Eq:seek-forecast}
\end{equation}
During the analysis-forecast cycles, each error mode evolves over time. The analysis error covariance is evaluated directly at each
analysis step as follows:
\begin{equation}
\textbf{P}^a_k = \textbf{S}^f_k [{\bf I} +(\textbf{H}_k \textbf{S}^f_k)^T 
\textbf{R}^{-1}_k (\textbf{H}_k \textbf{S}^f_k)]^{-1} \textbf{S}^{fT}_k \\
\label{Eq:decompanalysis}
\end{equation}
In this formula, the diagnostic of the forecast error modes depends on the formulation. 
With the ``fixed basis'' SEEK filter, the forecast error modes are equal to the analysis 
error modes at the previous cycle $k$:
$$ \textbf{S}^{f}_{k+1} = \textbf{S}^{a}_{k}$$
With the ``evolutive basis'' SEEK filter, the forecast error modes evolve with the fully non-linear model: 
$$ {[\textbf{S}^{f}_{k+1}]}_j = M {[\textbf{x}^a_k +
[\textbf{S}^{a}_{k}]}_j]  - M[\textbf{x}^a_k] \qquad\qquad j=1,\ldots, r$$
 This procedure 
makes it possible to follow the time evolution of the model variability, but increases the 
computational cost by a factor of $r$. In the present study, the implementation with a fixed basis is chosen. 
\subsection{The variational approach}
\label{SSect:mathodes-var}
As mentioned earlier,  for the variational approach a
reduced-order approximation of the Incremental 4D-Var algorithm (Courtier \textit{et al.}, 1994) has been used.
In this algorithm, we assimilate data available at different times $t_1$, ...,$t_N$, and the initial condition at 
time $t_0$ is controlled through an increment
$\delta x$. The following cost function must then 
be minimized:
\begin{eqnarray}
J(\delta \textbf{x})&=& \frac{1}{2} (\delta \textbf{x})^T\tbf{B}^{-1}\delta \textbf{x} + \nonumber \\
\frac{1}{2}&\displaystyle{\sum_{i=1}^{N}}&({\bf{H_{i}M_i}}\delta \textbf{x} 
-\bf{d}_i)^T R_{i}^{-1}({\bf{H_{i}M_i}}\delta \textbf{x} -\bf{d}_i)
\label{Eq:jtot}
\end{eqnarray}
where $\delta \tbf{x} = \tbf{x}(t_0) - \tbf{x}^b$ is the increment, and $\tbf{x}^b$  the first guess (or ``background'' value) for the model state at the initial time $t_0$. ${\bf M}_i$ 
is the tangent linear model between time $t_0$ and time $t_i$, $\tbf{H}_i$ is the
linearized observation operator at time $t_i$ and $\tbf{d}_i$ the
innovation vector $\bf{d}_i = \textbf{y}^o_i - {\bf{H_iM_i\textbf{x}^b}}$ ( $\textbf{y}^o_i$ is  the
observation vector  at time $t_i$). \par
Because of its size, the covariance matrix is never explicitly calculated in the full 4D-Var method. 
The \textbf{B} matrix is built as an operator composition in
order to represent error covariances, generally as gaussian-like functions (Weaver \textit{et al.}, 2001).\par
In the reduced-order approach, as proposed in section \ref{Sect:methods-subspace},  the increment $\delta \tbf{x} $ is looked for in a
low-dimension space spanned by the $r$ first EOFs: $\displaystyle{  \delta \tbf{x} = \sum_{j=1}^r w_j {\bf L}_j = {\bf L} {\bf w} }$, which results in the use of the low-rank covariance matrix ${\bf B}_r = {\bf L}  {\bf \Lambda}_r {\bf L}^T$. Formally, the same cost function (Eq. \ref{Eq:jtot}) must
be minimized (only the expression of \tbf{B} and $\delta \tbf{x}$ change), but the
minimization phase is performed on a very limited
number of coefficients $w_1, \ldots, w_r$.
When this reduced-order approach is compared to the full 4D-Var algorithm using the same twin-experiment framework as in the present paper,
it is found that only 10-15
iterations are needed to reach the minimum of the cost function while
almost 40 (and often more) are necessary with the full 4D-Var (Robert \textit{et al.}, 2005). 
For the assimilation of real data, however, designing a relevant 
reduced basis becomes a challenge, because
the model is no longer perfect. 
\subsection{The hybrid method}
\label{SSect:methods-hybrid}
These two reduced-order methods, SEEK filter and R-4D-Var, present several similarities. In particular, the choice of the initial 
error subspace can be exactly the same). 
However, intrinsic differences remain. For example, for the SEEK filter, the observations are
unrealistically co-located in time according to the analysis window (they are typically gathered every 10 days), unlike the 4D-Var in
which the observations are correctly distributed over time throughout the assimilation window (typically one month here). 
A second fundamental difference is that, in the SEEK filter, the error subspace
evolves in time at every analysis step, which makes it possible to follow the evolution of the error. 
In the R-4D-Var approach, the initial subspace generally remains constant during the assimilation period, 
even if this period is divided into successive time windows (e.g. one month) for the validity of the tangent linear approximation. \par
In an attempt to combine the best features of  both these methods, Veers\'e \textit{et al.}
(2000) proposed a hybrid algorithm using
the 4D-Var and the
 SEEK smoother. This method, developed only from a theoretical point of view, has never been
implemented in a real numerical configuration.\par
Although the theoretical context is slightly different here, since Veers\'e \textit{et al.}
(2000)  used a SEEK smoother instead of a SEEK filter, we retained the idea of making  the covariance matrix ${\bf B}$ of the R-4D-Var evolve in time thanks to the SEEK filter. The following hybrid algorithm can thus be proposed:
\begin{itemize}
\item Initialize ${\bf B} = {\bf P}_0^f$ using the $r$ first vectors provided by an EOF analysis
\item Perform  R-4D-Var and SEEK filter assimilations on successive time windows. In the present implementation, these windows are one month long. R-4D-Var processes this window in one go. For the SEEK filter, since the observations are artificially gathered every ten days, three analysis steps are performed during each window.
\item At the end of each window, ${\bf B}$ is updated in the R-4D-Var by the new value of ${\bf P}^f$ provided by the SEEK filter, and the state vector $\textbf{x}^f$ of the SEEK filter is reinitialized using the final state provided by the R-4D-Var at the end of the window. 
\end{itemize}
 The cost of this algorithm is the sum of the costs of both methods.
\section{Experiments}
\label{Sect:config-exps}
\subsection{Configuration}
\label{SSect:exps-config}
As mentioned previously, we compared  the different assimilation methods in a unique
 configuration in the tropical Pacific Ocean. The
general ocean circulation in this
area is weakly non-linear, which was seen as a convenient property for
conducting a first comparison of the methods.\par
\begin{figure}[H]
\begin{center}
\includegraphics[width=8cm, height=4cm, angle=0]{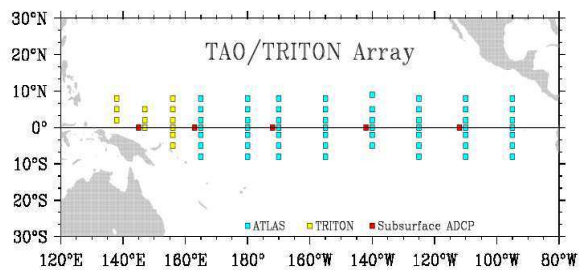}
\caption{TAO/TRITON array (http://www.pmel.noaa.gov)}
\label{Fig:TAO}
\end{center}
\end{figure}

The numerical model used in the experiments is the OPA model (Madec
\textit{et al.}, 1998), 
in the so called OPA-TDH configuration (Vialard \textit{et al.}, 2003). The extent of the domain is shown in Fig. \ref{Fig:TAO}.
The horizontal grid of the model is $1^\circ$ in longitude and $0.5^\circ$ in latitude at the equator,
stretched to reach $2^\circ$ at the northern and southern limits of the domain.
The vertical grid is composed of 25 levels, spaced at intervals ranging from 5 m at the surface to 1000 m for
the deepest levels.\par
Following Weaver \textit{et al.} (2003) and 
Vialard \textit{et al.} (2003),  
the year 1993 was chosen as our simulation period. 
During this year, the circulation of the tropical Pacific Ocean was 
marked by the weak influence of the last \nino{} event (the last big
one had occurred in 1982-1983, the ones in 1986-1987 and 
1991-1992 had been weaker and the next
big one would begin only in 1997). The year 1993 can therefore be seen as a ``normal" year from a dynamical point of view and was thus
suitable for conducting a comparison of the two different approaches. All experiments last one year, beginning on January 1, 1993.
The winds used to force the numerical model were based on both satellite ERS measurements and
in-situ TAO winds (Menkes \textit{et al.}, 1998). The atmospheric
heat fluxes came from ECMWF data files (ERA 40).
\subsection{Assimilation Experiments}
\label{Sect:exps}
All the experiments discussed here are conducted using the twin experiment framework.  The initial true state is obtained from a previous simulation over year 1992 and is used to generate a reference one-year free run considered as the truth. This
solution is then sampled to generate simulated temperature data. The
distribution of these simulated data is chosen  as close as possible to the distribution of
the real TAO/TRITON array (Fig. \ref{Fig:TAO}) and XBT profiles. Temperature is sampled from the surface down to a depth of 500 meters every six hours.
A gaussian noise is added to the simulated
observations with a standard error set at $\sigma_T = 0.5^\circ C$. \par
For the computation of the EOFs, the model state ${\bf x}$ consists of 4 variables: temperature, 
salinity and the two horizontal components of
velocity.  A free run experiment trajectory is sampled over one year,
using  a 2-day periodicity to build the covariance matrix. A large part of the total
variance is represented by a few EOFs: 80\% for the first 13 EOFs, 92\% for the first 30 EOFs.\par 
Since the initial state used in the assimilation experiments is not the correct one, we want to control an error on the initial condition through data assimilation.
This error is more or less corrected naturally by the free run in roughly six
months, thanks to the forcings. After six months, the error to be controlled is no longer an error on the initial condition but
mainly concerns the non-linear dynamics of the model.\par
In the R-4D-Var experiment, the error on the initial condition is introduced via the background.  For the initial background state, we use the solution of the reference simulation on April 1 (\textit{i.e.} 3 months later than the true state).   For the SEEK filter, according to the theory, the initial state is a mean state
calculated from the free run for the year 1993. In both cases, the initial error between the assimilation experiment and the reference run is large enough to make the correction, provided by the data
assimilation method, significant. A third assimilation experiment using the hybrid method presented previously was also performed. Finally, 
an additional fourth simulation, without data assimilation and using the same "false" initial condition as in the R-4D-Var experiment was also performed in order to help quantify the efficiency of the assimilation.
\par
Note that assimilating only temperature data at TAO points represents a departure from most previous SEEK 
filter analysis experiments. Surface fields like SSH are also usually assimilated, 
considerably  helping to constrain the solution and acting more specifically on the dynamics. Moreover, it should be noted that the limitation of the R-4D-Var method arising from the fact that the model is a strong constraint does not play a role here since the model is supposed to be perfect in these twin experiments.\par
For the variational approach we used the OPAVAR package, developed and
validated by Weaver \textit{et al.} (2003) and Vialard \textit{et al.} (2003), and for the SEEK filter experiments, the SESAM
package (Testut \textit{et al.}, 2001).
\section{Results}
\label{Sect:results}
The results shown below are presented for two different periods, from January to June 1993 and from July  to December 1993. There are  two main reasons for this: (i) in such a quasi-linear model and twin-experiment framework, the model is naturally restored from the erroneous initial condition over a time scale of some months, (ii) physically, June is also the time of the onset of the non-linear Tropical Instability Waves (TIWs) in the eastern tropical Pacific Ocean. Schematically, the first six-month period concerns the control of the error on the initial condition, while the next six-month period concerns the control of the non-linear dynamics in the system. In the latter period, the intensity of TIWs begins to increase, a development that appears to have  considerable influence on data assimilation. 
\subsection{January-June 1993}
\label{Sect:results-exps1}
\begin{figure}[H]
\centering
\subfigure[Temperature]{\includegraphics[width=5cm, height=4cm, angle=0]{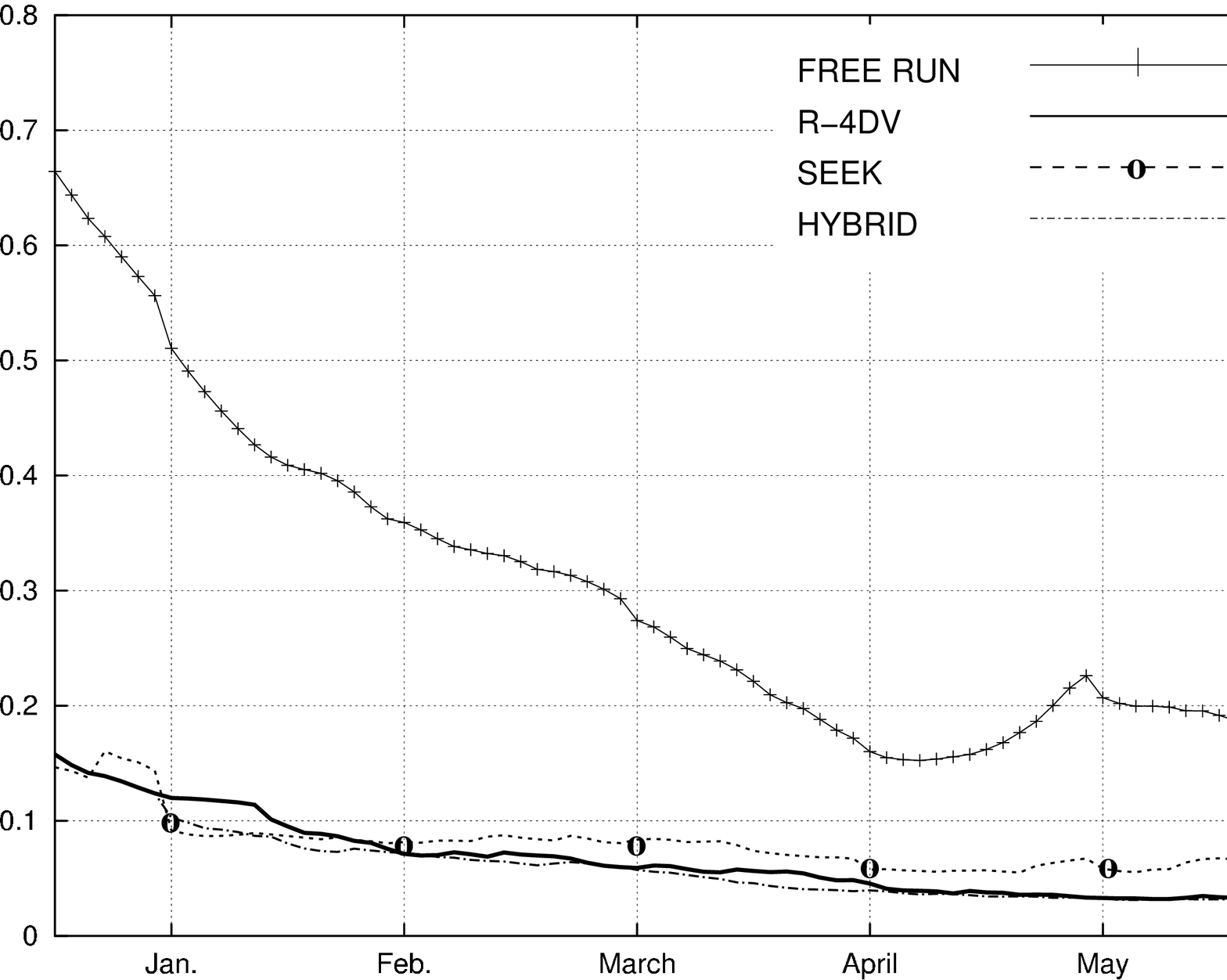}}
\subfigure[Salinity]{\includegraphics[width=5cm, height=4cm, angle=0]{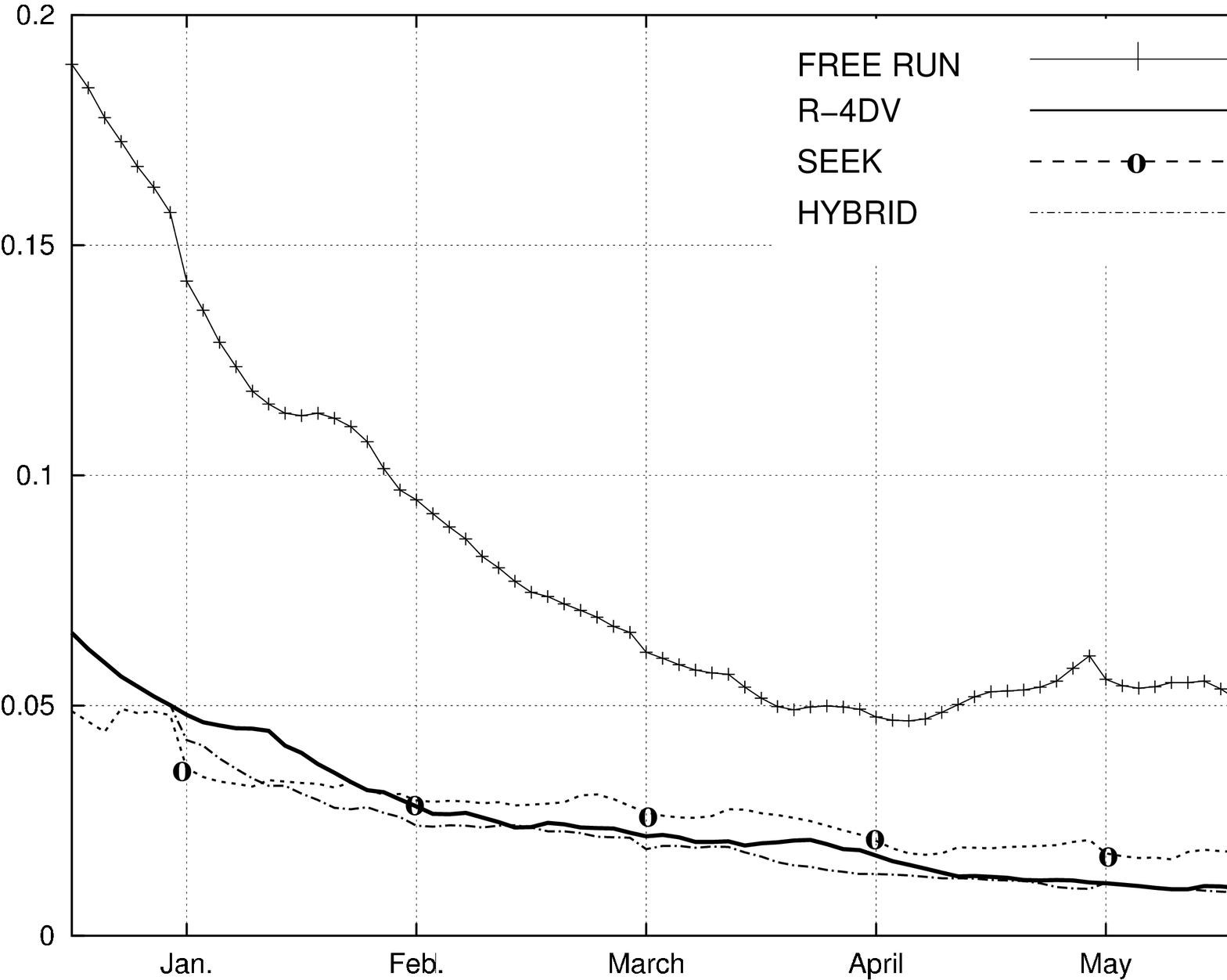}}
\subfigure[Velocity \textbf{u}]{\includegraphics[width=5cm, height=4cm, angle=0]{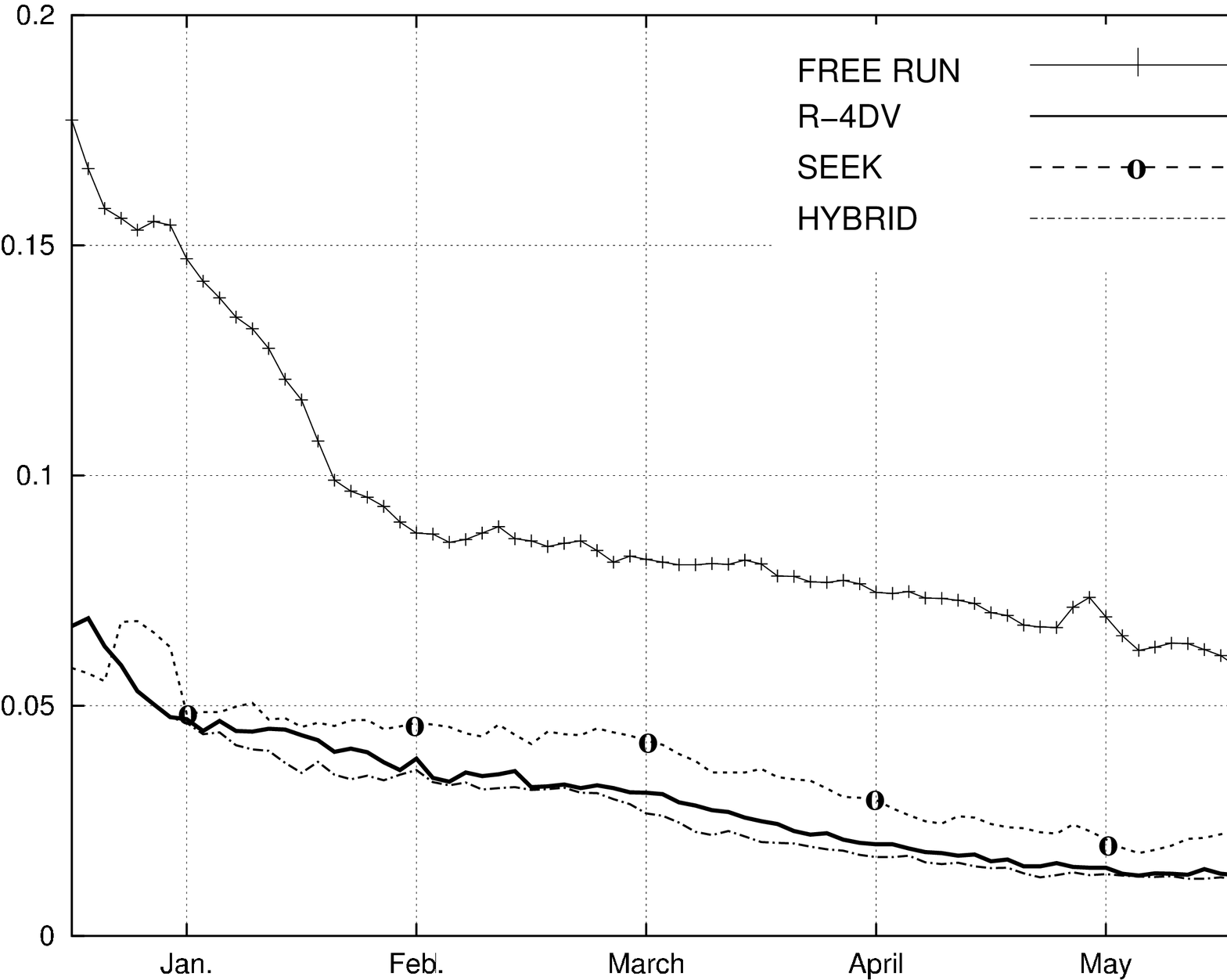}}
\subfigure[Velocity \textbf{v}]{\includegraphics[width=5cm, height=4cm, angle=0]{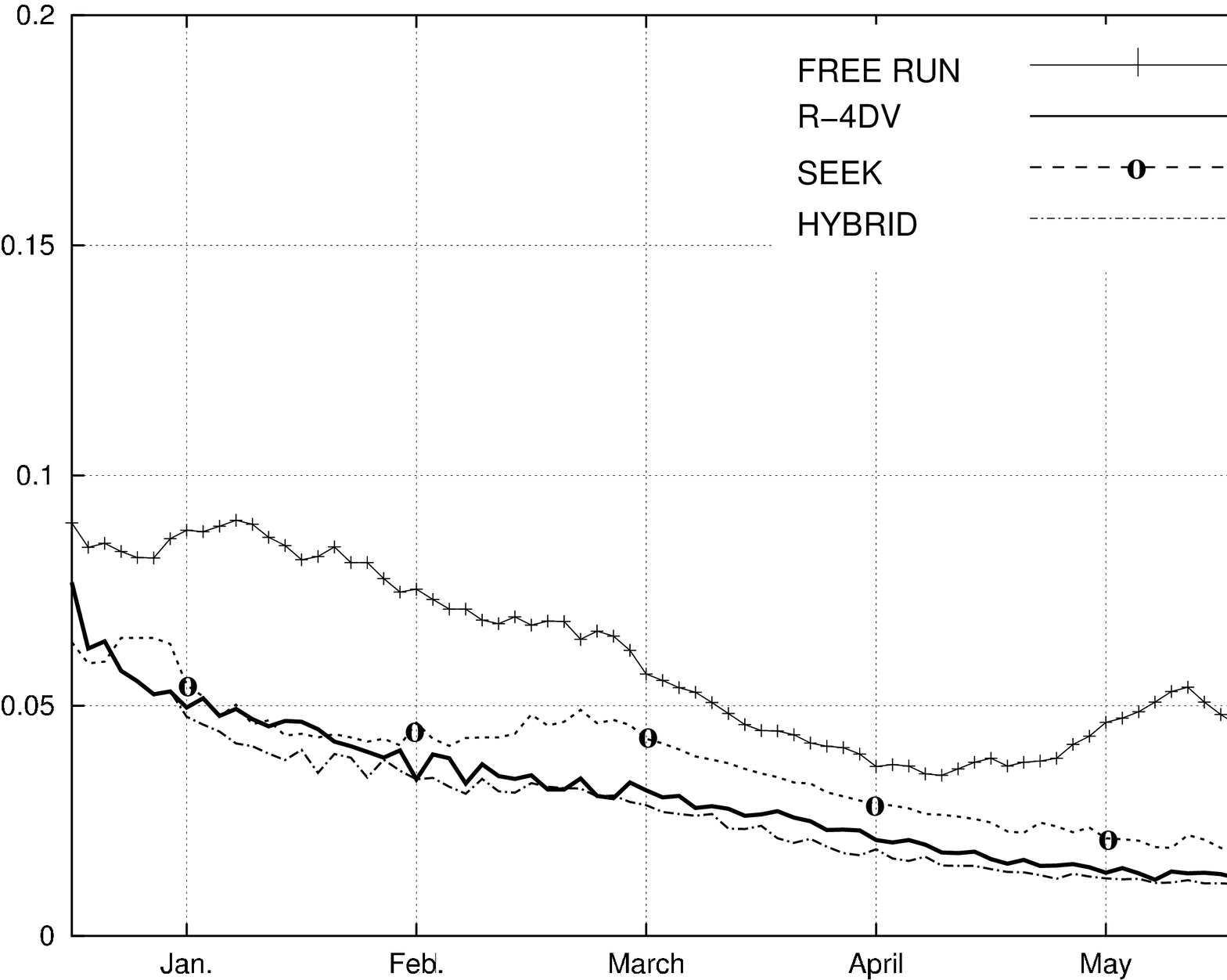}}
\caption{Absolute Rms error obtained by each method during the
first 6 months of simulation, at 15m. depth. Solid line with +: free run, solid line: R-4D-Var,
dashed line with o: SEEK filter and dashed-dotted line: hybrid method.}
\label{Fig:RMS-6M1-K2}
\end{figure}

\begin{figure}[H]
\centering
\subfigure[Velocity \textbf{u}]{\includegraphics[width=5cm, height=4cm,
angle=0]{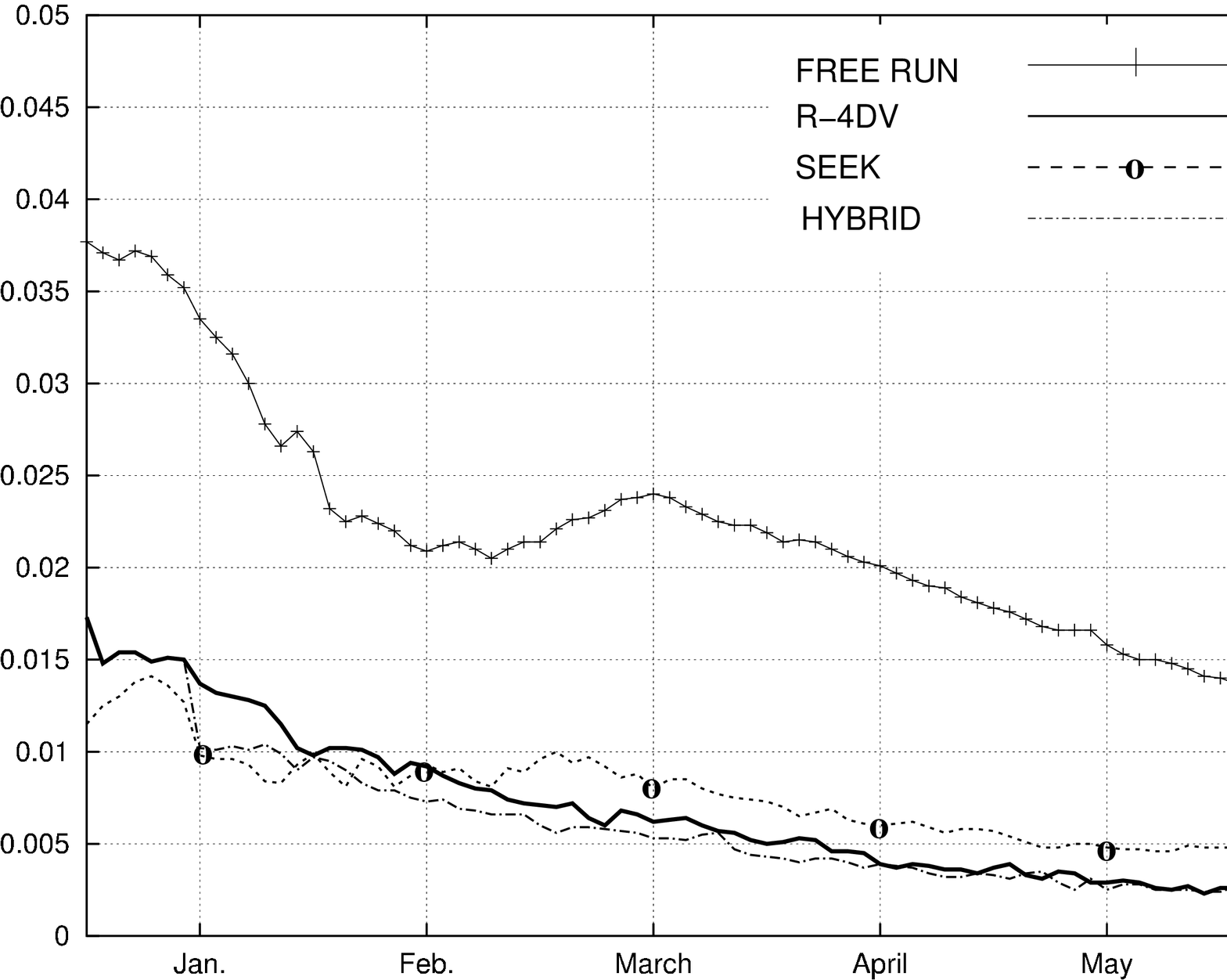}}
\subfigure[Velocity \textbf{v}]{\includegraphics[width=5cm, height=4cm,
angle=0]{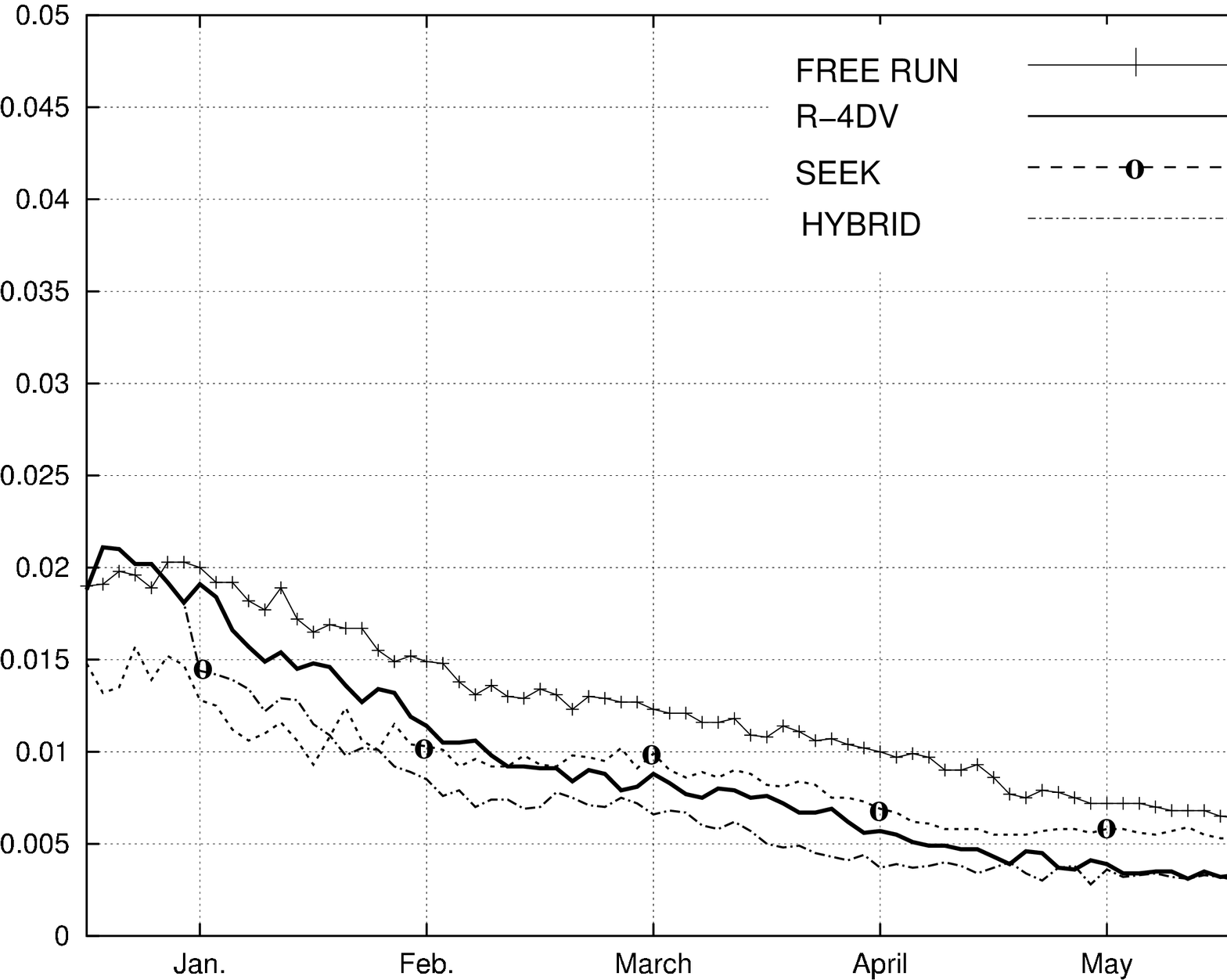}}
\caption{Rms error obtained by each method during the first 6 months
of simulation, below the observations, at 750 m. depth. Solid line with +: free run, solid line: R-4D-Var,
dashed line with o: SEEK filter and dashed-dotted line: hybrid method.}
\label{Fig:RMS-6M1-K20}
\end{figure}

In this first time period, the error on the initial condition is quite large, since the forcings of the model have not had time to correct it, so that most of the work done by data assimilation is to  control the error on the initial condition.
Moreover, the dynamics is quite stable and the EOFs represent model variability  perfectly. 
In this case, and as would be expected from the purely linear and optimal
context, it can be observed in Fig. \ref{Fig:RMS-6M1-K2} that the assimilation methods are almost equivalent. We can see  that the correction is substantial and that all 
methods provide roughly the same solution. 
The three algorithms work well, not only in terms of temperature misfit in the area 
of observations but also at greater depths and for the other variables, thanks to the
multivariate nature of the EOFs. This can be seen for example in Fig. \ref{Fig:RMS-6M1-K20}.\par
Concerning the hybrid algorithm, 
a slightly lower level of error is obtained as shown in Fig. \ref{Fig:RMS-6M1-K2}, for example
for $(u,v)$ variables. 
The diagnostic of the analysis and forecast errors performed by the SEEK filter is
 correct with regard to the dynamics.
The solution is thus very good. Finally, this hybrid method succeeds in combining two intrinsic 
aspects of the reduced-order methods, which leads to slightly improved results. The fact that the improvement is not more significant is because both  the SEEK and R-4D-Var methods already obtain excellent results in decreasing the error.
\subsection{July-December 1993}
\label{Sect:results-exps2}
The second period starts in July, when the intensity of TIWs increases significantly. 
 The dynamics of these waves is non-linear. Since the first TAO point is quite distant from the eastern coast, the TIWs rise before the first data point can see the change.
In this case, it can take a significant time (even more than 10 days, which is the duration of the SEEK cycle) before the easternmost 
points of the TAO array register these changes. \par
\begin{figure}[H]
\centering
\subfigure[End of June]{\includegraphics[width=7cm, height=5cm, angle=0]{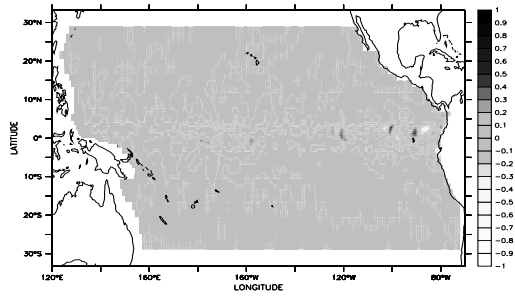}}
\subfigure[Beginning of July]{\includegraphics[width=7cm, height=5cm, angle=0]{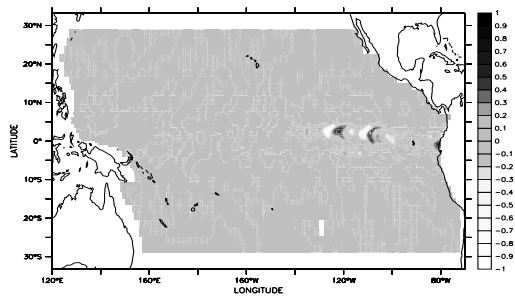}}
\caption{Difference between temperature field at 15 m. depth of the reference simulation and of the SEEK filter
simulation.}
\label{Fig:RMS-SPAT-K2}
\end{figure}

In addition, the error due to the initial condition is very weak,
since the dynamics of the model has already naturally corrected this error. Thus, the major part of the remaining error
is driven by the non-linear dynamics of the ocean. As we can see in Fig. \ref{Fig:RMS-SPAT-K2}, 
the difference in temperature between the reference simulation and the SEEK filter simulation is mainly located in the
eastern basin, near the equator.\par
\begin{figure}[H]
\centering
\subfigure[Temperature]{\includegraphics[width=5cm, height=4cm]{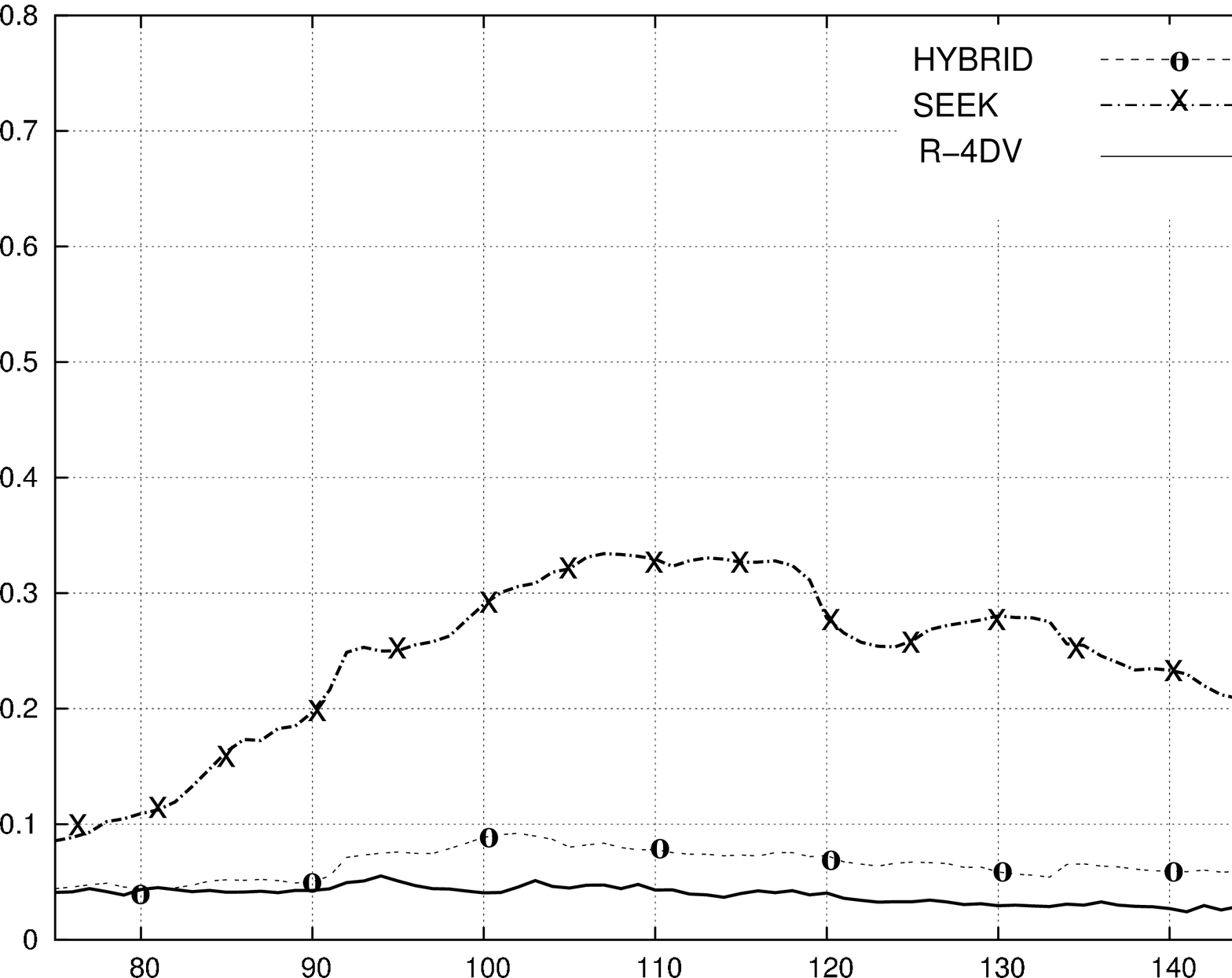}}
\subfigure[Salinity]{\includegraphics[width=5cm, height=4cm, angle=-0]{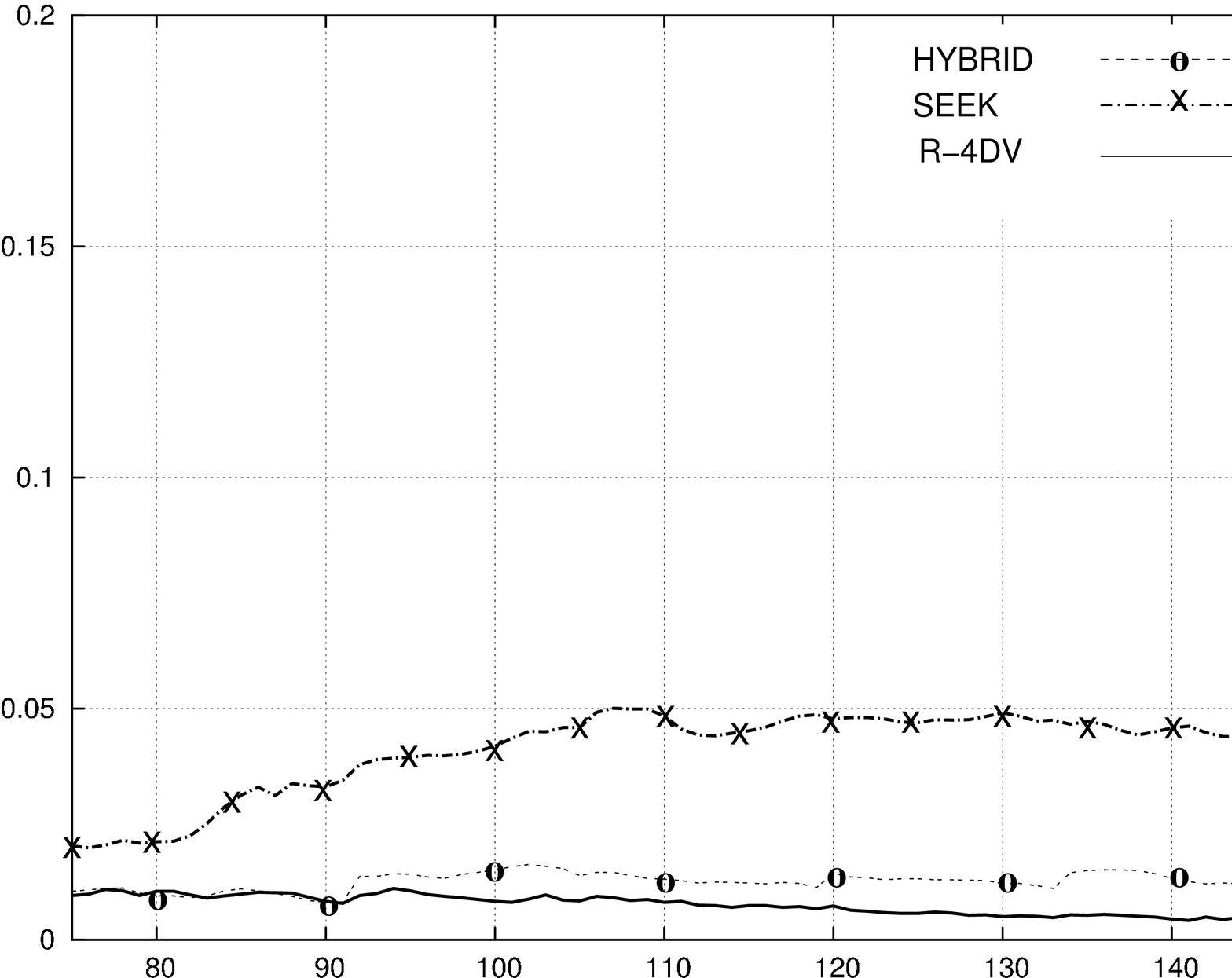}}
\subfigure[Velocity \textbf{u}]{\includegraphics[width=5cm, height=4cm, angle=-0]{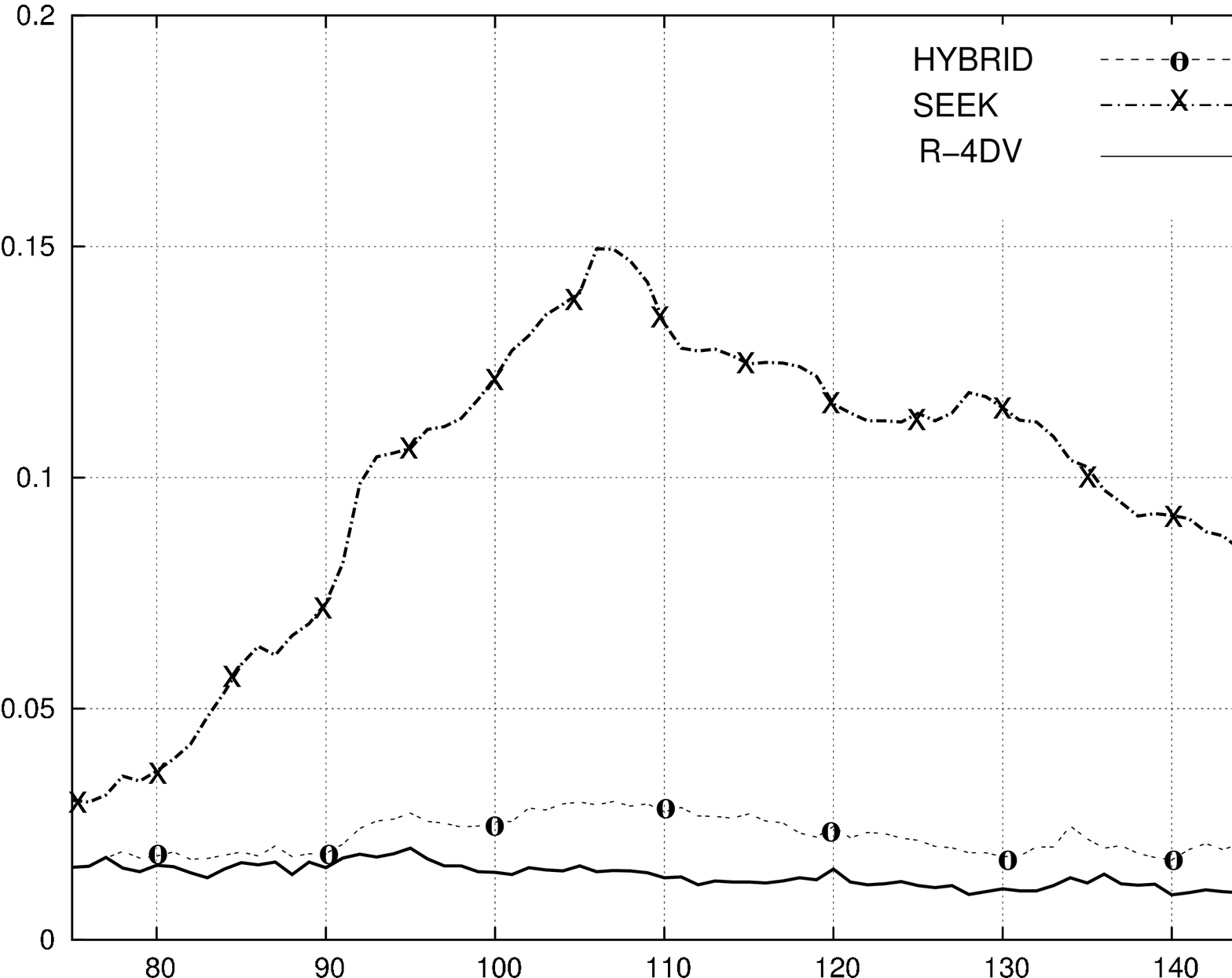}}
\subfigure[Velocity \textbf{v}]{\includegraphics[width=5cm, height=4cm, angle=-0]{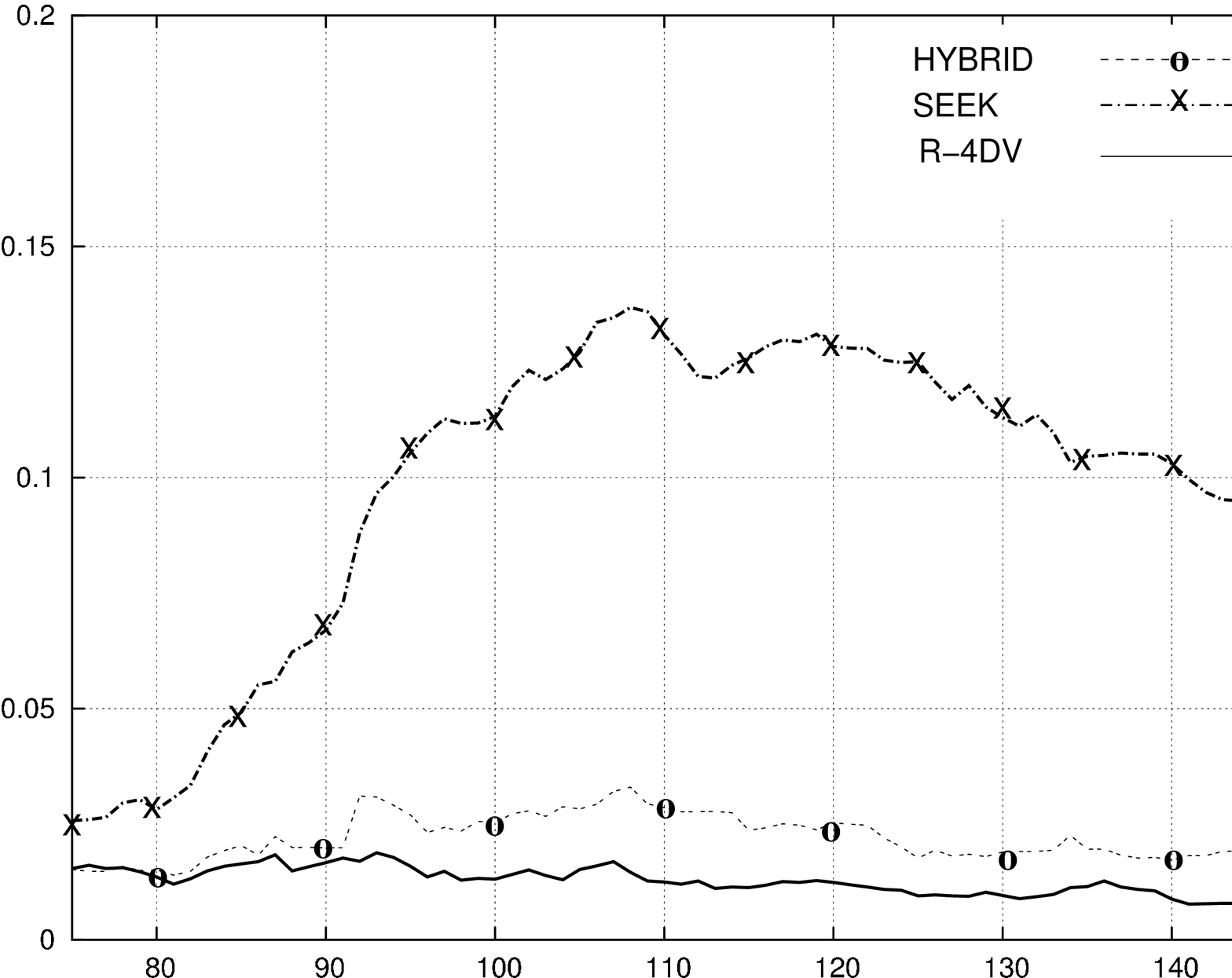}}
\caption{Rms error the last 6 months of simulation, at 15 m. depth, in the eastern part of the basin. Solid line: R-4D-Var,
dashed line with o: hybrid method and dashed-dotted line with x: SEEK filter.}
\label{Fig:RMS-z2-6M2-K2}
\end{figure}
\begin{figure}[H]
\centering
\subfigure[Temperature]{\includegraphics[width=5cm, height=4cm, angle=0]{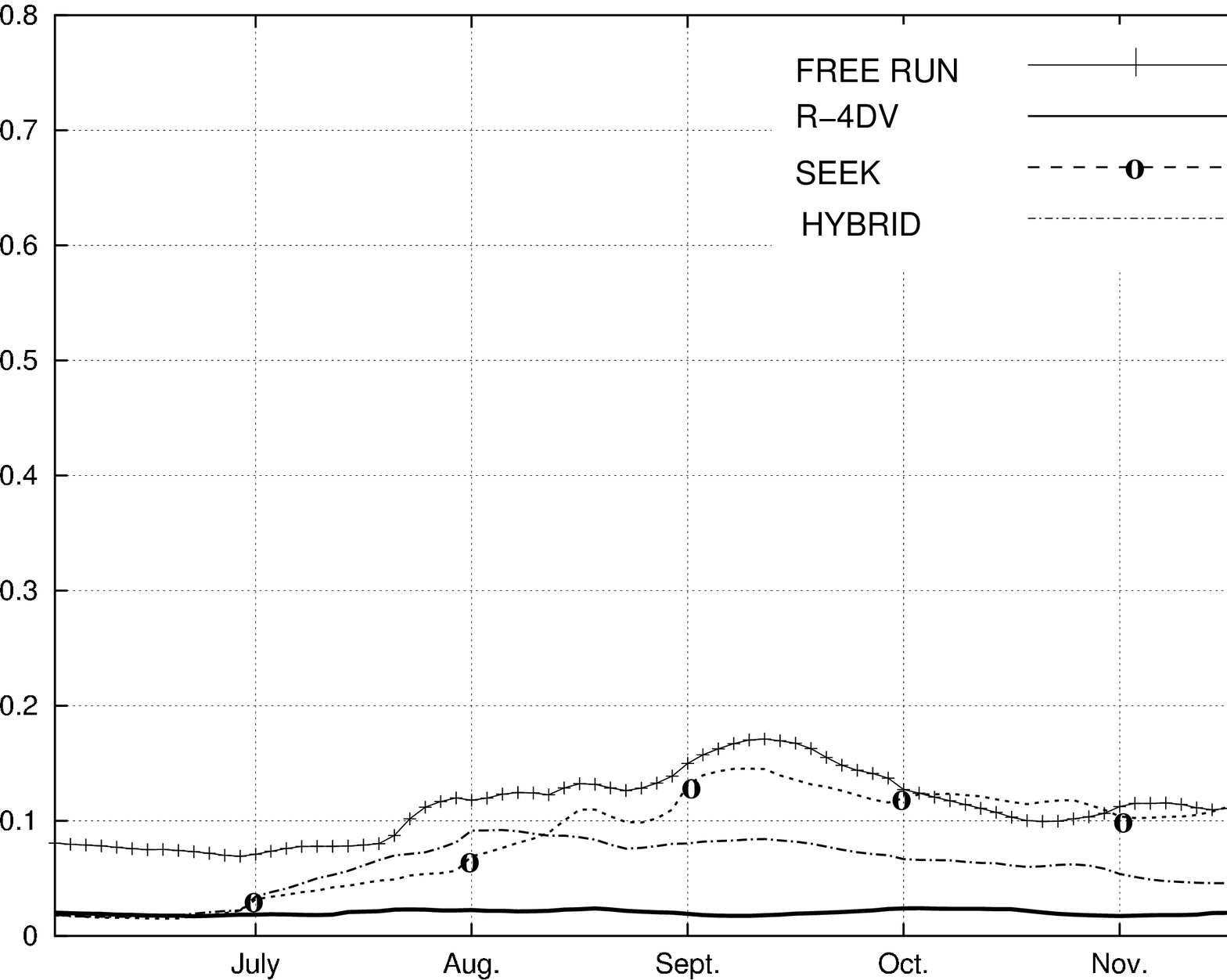}}
\subfigure[Salinity]{\includegraphics[width=5cm, height=4cm, angle=0]{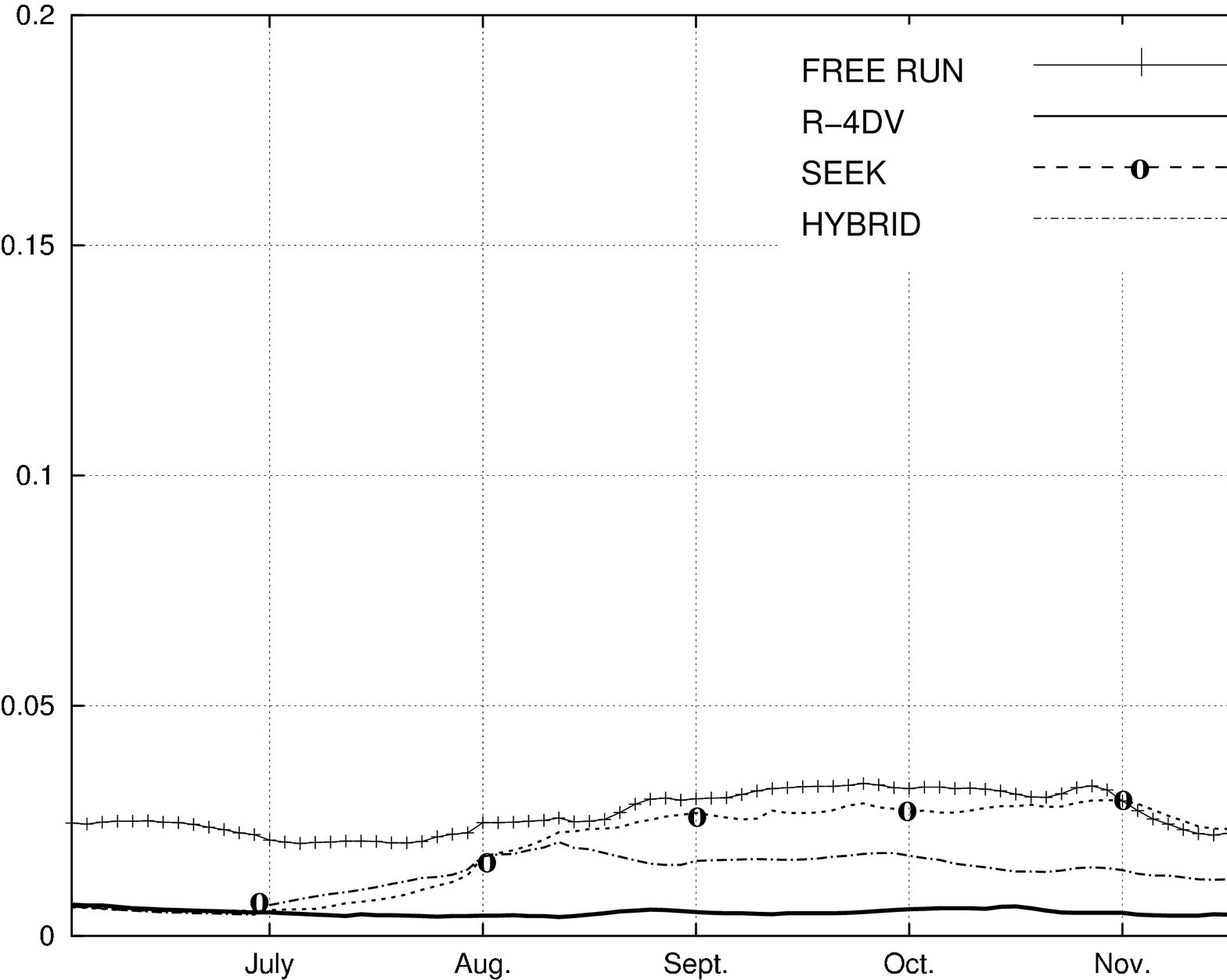}}
\subfigure[Velocity \textbf{u}]{\includegraphics[width=5cm, height=4cm, angle=0]{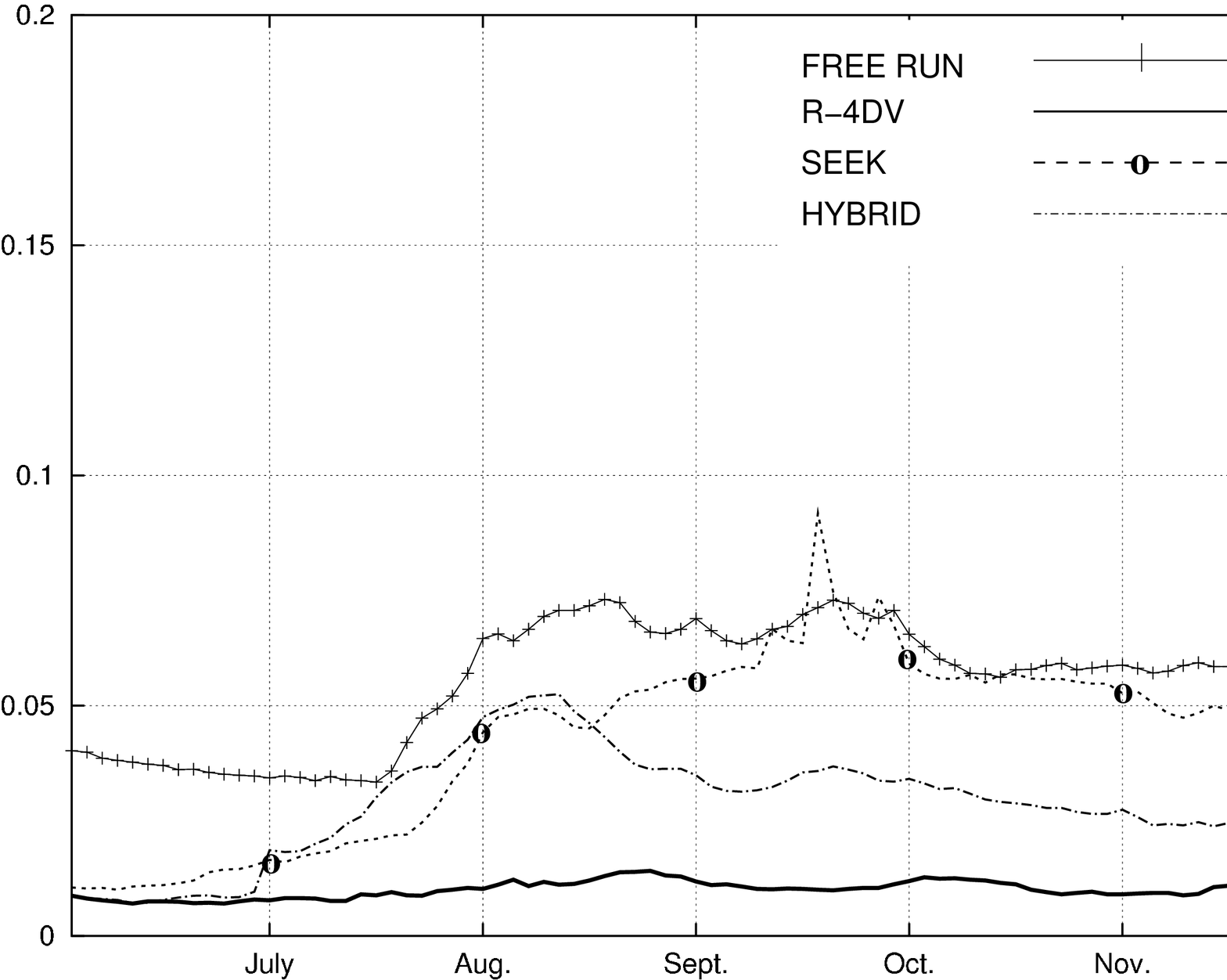}}
\subfigure[Velocity \textbf{v}]{\includegraphics[width=5cm, height=4cm, angle=0]{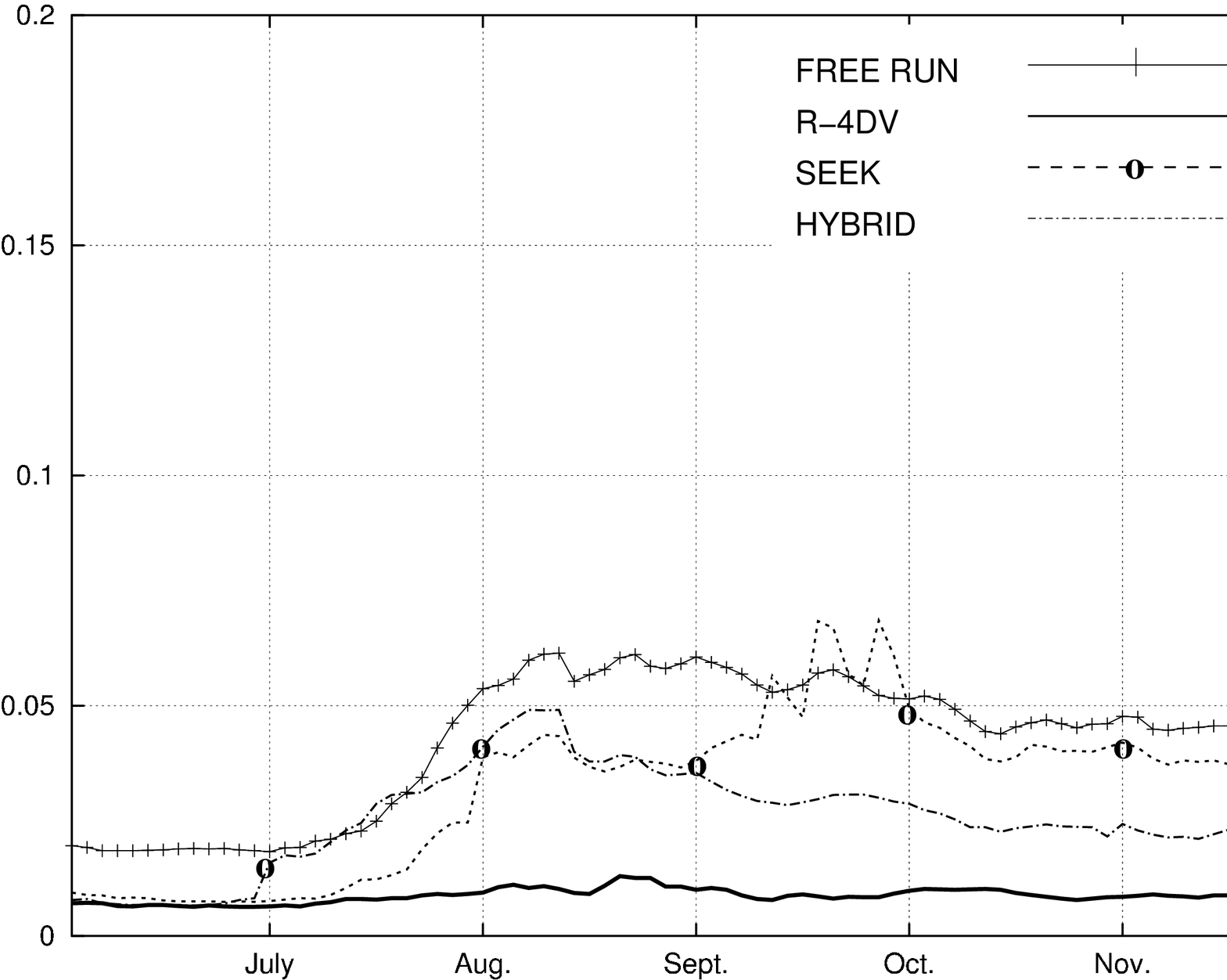}}
\caption{Rms error the last 6 months of simulation, at 15 m. depth, in the western part of the basin. Solid line with +: free run, solid line: R-4D-Var,
dashed line with o: SEEK filter and dashed-dotted line: hybrid method.}
\label{Fig:RMS-z1-6M2-K2}
\end{figure}

An important difference between the algorithms is their sequential versus global processing of the observations.  Since the observations 
available during the whole assimilation window are taken into account in the R-4D-Var method, this approach can anticipate the propagation 
of a physical phenomenon at this time scale. For
example, the Tropical Instability Waves (TIWs) rise in the eastern part of the basin and propagate along 
the equator in roughly one month. When they become more intense in early July, the variational 
system takes into account observations of these waves 
in the analysis conducted before their actual 
occurrence. This is not the case in the SEEK filter because the analysis 
at the same time takes into account only past
observations. Consequently, there is a time lag with this approach between the occurrence of the physical phenomenon (the intensification) and its integration into the analysis. 
This explains the differences observed in Fig. \ref{Fig:RMS-z2-6M2-K2} which concerns the eastern part of the domain. The comparison is
 almost the same in the western part (Fig. \ref{Fig:RMS-z1-6M2-K2}), but with a lower rms error level, probably because more observations are available.\par
In that second period, the hybrid method continues to draw advantages from the quality of the 
analysis of the R-4D-Var. However, since the evolution of its covariance matrix provided by the SEEK filter 
is less accurate, the hybrid method do not succeed in that case in providing better results than
the R-4D-Var (see Fig \ref{Fig:RMS-z1-6M2-K2} and \ref{Fig:RMS-z2-6M2-K2}).
\section{Conclusions}
\label{Sect:conclu}
This paper presents the results of a first comparison of three reduced-order data assimilation methods implemented  in a model of the tropical Pacific Ocean. The
first two methods, the SEEK 
filter and the reduced-order 4D-VAR, are respectively derived from the sequential and  the variational approaches. The third method, combining features of the SEEK filter
and the R-4D-VAR, is a hybrid version of the first two methods.
To our knowledge, the present study is probably one of the first side-by-side
implementations and comparisons  of these techniques ever made in a
realistic context. Investigations are exploratory in nature due to the complexity of the methodology 
and more especially because simulations have been carried out in a twin-experiment framework
where no model error is present and data are simulated. In addition, only one year of comparative 
simulations is performed. However, we believe that the first results
provide useful insights:
\begin{itemize}
\item In a quasi-linear regime, as expected from linear theory,  the three methods provide rather 
similar results in reducing the initial condition system error. The hybrid method provides slightly 
better results, which would mean, as expected, that combining the evolution of error covariance matrices and the variational analysis is, at the very least, feasible and potentially fruitful (as soon as further tuning is done).
\item In a regime where strong nonlinearities develop at regional scales 
(corresponding to the onset of Tropical Instability Waves within the eastern tropical Pacific), 
the 4D-VAR succeeds in keeping the error to a low level whereas the SEEK filter, due to its sequential nature, fails to fully 
control the increasing instability using the data available. The
hybrid method follows the divergent nature of the SEEK filter in 
the first stages but, after several weeks, resumes a more convergent path.
\end{itemize}
The twin-experiment set-up entails obvious limits and may influence
the conclusions. 
The``no model error" assumption may favor the variational solution since the twin 
experiments are exactly within the ``strong constraint" variational  framework. 
It also favors  the performances of all reduced-order methods since the reduced basis 
can be built from a perfect reference simulation. With statistical low-cost 
methods like the SEEK filter, the amount and the nature of assimilated data are a key factor. The data used in the present study
mimic the real data that are acquired from the TAO array in the Pacific, but this array 
poorly samples the eastern Pacific Ocean. It is likely that with the addition of 
some higher frequency complementary data (such as altimetric data), the SEEK filter 
would behave more satisfactorily (see, for example, Castruccio \textit{et al.}, 2006). Preliminary experiments were also performed with 
real TAO data, thus including model  errors. In this case, first results 
seem to indicate a much more balanced behavior between the SEEK filter and the reduced 4D-VAR.\par
\vspace*{5mm}
{\bf Acknowledgments}\par
J.-M. Brankart and A. Weaver are gratefully acknowledged for supplying 
respectively the SESAM and OPAVAR packages and for providing
support in using these tools.
 This work has been supported by CNES and the 
MERCATOR project. Idopt is a joint
CNRS-INPG-INRIA-UJF research project.\vspace*{3 mm}\\ 
\nocite{*}
\bibliographystyle{plain}
\bibliography{./Robert-ODYN}
\end{document}